\documentclass[aps,twocolumn,showpacs,prb]{revtex4-1}

\usepackage{color,vmargin,epsfig,rotating,graphicx,subfigure}
\usepackage{amsmath,amssymb,amscd}
\usepackage[latin1]{inputenc}
\usepackage[english]{babel}
\usepackage{dsfont}
\usepackage{textcomp}

\setpapersize{A4}

\newcommand{\ri}{{\rm i}}
\newcommand{\re}{{\rm e}}

\renewcommand{\vec}{\boldsymbol}

\begin{document}

\title{Plasmons in disordered nanoparticle chains: Localization and Transport} 

\author{Felix R\"uting }
\email[Email:]{rueting@theorie.physik.uni-oldenburg.de}
\altaffiliation[Current address: ]{Departamento de F{\' i}sica de la
  Materia Condensada, Universidad Aut{\'o}noma de Madrid.}
\affiliation{Institut f\"ur Physik, Carl von Ossietzky Universit\"at,
    	D-26111 Oldenburg, Germany}

\date{\today}

\begin{abstract}
Disorder-induced effects on plasmon coupling in chains of metallic
nanoparticles are studied within a dipole model, by considering two
types of disorder: fluctuations of the particles' shapes and fluctuations
of their positions. Typical localization effects are found both in the
eigenmodes and in the transport behavior of the system, and an
estimate of the localization length is made. It is argued that chains
with deliberately introduced disorder constitute promising systems for
studying localization effects of electromagnetic waves at optical
frequencies under well controllable and manipulable conditions.
\end{abstract}




\maketitle

\section{Introduction}
Disorder-induced localization was studied in detail for the
wavefunctions of electrons by P.~W. Anderson~\cite{Anderson58}. This
localization effect rests on the interference of multiply scattered
waves. Therefore, a similar effect is also present for electromagnetic
waves~\cite{Anderson85, *Wolf85, *Arya86, *John87, *Drake89,
  Wiersma97}, with the difference that there is no interaction between
the photons and, hence, localization does not have to compete against
interaction effects. In general, in three dimensions Anderson-localization of
waves is observable if the density of scatterers exceeds a critical
value, as expressed by the Ioffe-Regel
criterion~\cite{Ioffe60}: $kl\lesssim 1$, with $k$ denoting
the wavenumber, and $l$ the mean free-path length. Hence, for
electromagnetic waves in the visible spectrum (with $k$ on the order
of $10^7\, \text{m}^{-1}$) the mean free-path length has to be on the
order of a few nanometers. This can be achieved, for example, with
semi-conductor powders~\cite{Wiersma97}.

In the present paper, localization of an electromagnetic excitation
in disordered chains of metallic nanoparticles is studied. In the
context of integrated optics and plasmonic wave\-guides, the response of
nanoparticle chains without sizeable disorder to an optical excitation
with a frequency close to the plasmon resonance frequency of a single
chain member has already been investigated in detail both
experimentally~\cite{Maier03, Koenderink07,*deWaele07} and
theoretically~\cite{Quinten98, Brongersma00, Maier03a,
  Citrin04,Weber04,Citrin05,Hernandez05,Citrin06,Alu06,Alu10}. In
these studies it has mostly been assumed that the nanoparticles are
perfect spheres, and it has been shown that such ideal chains offer
interesting perspectives for sub-wavelength wave\-guiding.

Recently, Al{\'u} and Engheta have made an important step by studying
the effect of small random uncontrollable disorder on the guidance
properties of such chains, within first order perturbation
theory~\cite{Alu10}. These authors were able to quantify the resulting
additional radiation losses for the guided mode. Here, a further step
is taken by evaluating even strong disorder, and making the connection
to localization physics.

In a wider context, localization of surface plasmons has already been
reported experimentally for semicontinuous metal
films~\cite{Gresillon99}, and studied theoretically, e.\ g., for
randomly distributed metal particles~\cite{Arya86}. Also the
celebrated surface enhanced Raman spectroscopy (SERS) is basically
enabled by the formation of ``hot spots'' near a rough
surface~\cite{Kneipp97, *Nie97,*Kneipp02}. These hot spots are regions
with an extremely enhanced near-field, and result from the
localization of surface plasmons.

The disordered nanoparticle chains studied in this contribution offer
an alternative approach to such phenomena. When the individual
nanoparticles are subjected to shape fluctuations, and randomly
oriented, the polarization tensors of the chain elements have, in
general, three distinct principal axes, which are not aligned within
the chain. As will been shown by studying their optical properties,
such chains exhibit typical signatures of localized modes, when
excited at frequencies close to the plasmon-resonance frequency of a
single particle.  In contrast to systems with truly random disorder,
such systems may allow one to introduce specifically engineered types
of disorder, and therefore enable one to create hot spots at desired
positions, and with predefined properties.

The model used to describe the optical properties of the chains is
briefly introduced in Sec.~\ref{Sec:Model}. In Sec.~\ref{Sec:Results}
numerical results for two types of disorder are shown and some
concluding remarks are given in Sec.~\ref{Sec:Conclusions}.

\section{Model}
\label{Sec:Model}
The present theoretical study is based on a model of electromagnetic
transport through a chain of metallic nanoparticles which regards the
chain as a linear array of point
dipoles~\cite{Brongersma00,Maier03a,Citrin04,Weber04,Citrin05,Hernandez05,Citrin06}.
This approximation is viable as long as the individual chain members
are small compared to the relevant wavelengths, so that their optical
response can be described in the quasi-static approximation. Moreover,
it is required that the distance between the chain elements is large
compared to their linear size, so that higher multipoles are
negligible~\cite{ParkStroud04}. Experiments conducted in this
parameter regime are in good qualitative agreement with predictions
made by this model~\cite{Koenderink07,deWaele07}. This dipole model
was proposed in 2000 by Brongersma {\emph et al.}~\cite{Brongersma00},
and since then has been elaborated further in several studies. In
particular, Weber and Ford~\cite{Weber04} pointed out that while in
the parameter regime used, the quasi-static approximation is valid for
the response of a single chain member, for the coupling of the
particles the quasi-static model is not sufficient. Therefore, in
the following always the standard field produced by an oscillating
dipole $\vec{p}\re^{\ri \omega t}$ is used~\cite{Jackson}
\begin{align}
 \vec{E}=&\frac1{4\pi\epsilon_0}\left\{\frac{k^2}{r}(\vec{
   \hat{r} } \times \vec{p})\times
 \vec{\hat{r}}\right.\nonumber
\\+&\left.\left[3\vec{\hat{r}}(\vec{\hat{r}}\cdot\vec{p})-\vec{p}
   \right]\left(\frac1{r^3}-\frac{\ri k}{r^2} \right) \right\}e^{\ri
   kr}\label{Eq:dip}
\\=&\frac1{4\pi\epsilon_0}\left\{ \vec{p}\left[ \frac{\ri
     k}{r^2}-\frac1{r^3}+\frac{k^2}{r}\right]\right.\nonumber
\\ +&\left.\vec{\hat{r}}(\vec{\hat{r}\cdot\vec{p}})
 \left[\frac3{r^3}-\frac{3\ri k}{r^2}-\frac{k^2}{r} \right] \right\}e^{\ri
   kr},\nonumber
\end{align}
with the vector $\vec{r}$ pointing from the source to the observation
point ($\vec{\hat{r}}$ denotes the corresponding unit vector) and the
wave-number $k=\omega/c$. The time-dependence is suppressed in
Eq.~(\ref{Eq:dip}) and in the following (always assuming a harmonic
time-dependence $\re^{\ri \omega t}$).

 Assuming, to begin with, that all particles are spheres
surrounded by vacuum, the dipole moment $\vec{p}(\vec{r}_j)$ induced
in the $j$-th particle located at the position $\vec{r}_j$ is related
to the electric field $\vec{E}(\vec{r}_j)$ through its effective
polarizability $\alpha_{\rm eff}$:
\begin{equation}
\label{eq:polar}
\vec{p}(\vec{r}_j)=\epsilon_0\alpha_{\rm eff} \vec{E}(\vec{r}_j),
\end{equation}
with the vacuum permittivity $\epsilon_0$. The effective
polarizability, in its turn, is related to the quasi-static one
$\alpha$ via~\cite{Weber04,Wokaun82}
\begin{equation}
\label{eq:alpha}
\alpha_{\rm eff}=\frac{\alpha}{1-\frac{i \alpha k^3}{6 \pi}},
\end{equation}
with $k=\omega/c$ and $c$ denoting the vacuum speed of light. Here,
the radiation damping is included within the Abraham-Lorentz
approximation~\cite{Jackson}. Therefore, the model can be expected to
be quantitatively correct as long as radiation damping is a small
correction. Since the electric field at each chain site is the sum of
the applied external field and the fields produced by all other
dipoles, Eq.~(\ref{eq:polar}) describes a coupled system of equations
for the moments at the sites.

When replacing the spheres by arbitrarily deformed particles their
polarizabilities have to be described by tensors, instead of scalar
polarizabilities $\alpha$ and $\alpha_{\rm eff}$. Then the induced
dipoles and the local electric fields are no longer
parallel. Accordingly, the effective polarizabilities then are defined
by diagonalizing that tensor, and using Eq.~(\ref{eq:alpha}) for
each diagonal element. The tensorial character of the polarizabilities
implies that a decoupling of transversal and longitudinal modes is no
longer possible in chains of randomly oriented, distorted
particles. If a chain of ideal spheres is excited with a field
polarized perpendicular to the chain-axis, symmetry requires that also
all induced dipoles are oriented perpendicular to this axis
(transversal mode); similarly a parallel excitation leads to dipoles
oriented along the chain axis (longitudinal mode). While this
categorization is very helpful for modeling the perfect chain and also
enables one to derive some analytical results in this case, it is no
longer valid for a disordered chain. Of course, the discrimination of
transversal and longitudinal \emph{excitation} remains meaningful.

This set of equations (\ref{eq:polar}), augmented by tensorial
polarizabilities for each particle, allows one to investigate the
propagation of an optical excitation through a chain of disordered
metallic nanoparticles in detail. In the present study mainly chains of
randomly oriented ellipsoids are considered, and both eigenmodes and
transport properties are calculated. As a second type of disorder also
chains of perfect spheres but with variations of the particle
positions is used.

\section{Numerical Results}
\label{Sec:Results}
\begin{figure}[htb]
\epsfig{file=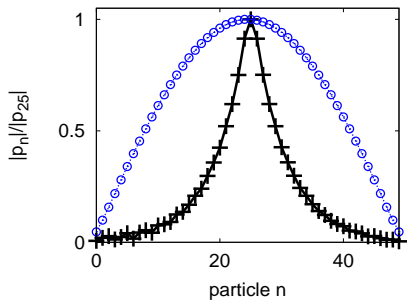, width=0.4\textwidth}
\caption{(Color online) Extended, standing-wave like mode in a chain
  of 50 silver nanoparticles consisting of perfect spheres (dashed
  line with circles) compared to a localized eigenmode which emerges
  if one single sphere is replaced by an ellipsoid (solid line with
  crosses). The resonance frequency of this mode is $\omega=(1.0912 -
  0.147296 i)\, \omega_{spp}$, where $\omega_{spp}=\omega_p/\sqrt{3}$
  is the resonance frequency of a sphere.}
\label{Fig:F_1}
\end{figure}

Let us start by investigating a chain of 50 ideal spheres, with a
center-to-center distance of $80\, \text{nm}$ and a sphere-radius of
$a_0=25\,\text{nm}$, assuming that the dielectric properties of the
particles are described by the Drude model
\begin{equation*}
\epsilon(\omega)=\epsilon_0 \left[ 1-\frac{\omega_p^2}{\omega^2+\ri
    \omega\tau^{-1}} \right]
\end{equation*}
with plasma frequency $\omega_p=1.4 \cdot 10^{16}\,\text{s}^{-1}$ and
relaxation time $\tau=1.27 \cdot 10^{-14}\, \text{s}$, as appropriate
for silver~\cite{Brongersma00}. In Fig.~\ref{Fig:F_1} an
eigenmode~\cite{footnote2} of this chain is depicted by plotting the
induced dipole moments $|p_n|=|\vec{p}(\vec{r}_n)|$; this mode can be
recognized as a standing wave~\cite{Weber04}. This changes drastically
when one individual sphere in the chain center is replaced by an
ellipsoid, here with semi-axes $0.98\,a_0$, $0.96\, a_0$, and $1.02\,
a_0$, representing an isolated defect. As shown in Fig.~\ref{Fig:F_1}
this defect leads to the emergence of an exponentially localized mode,
similar to the localization of electronic wavefunctions around
isolated lattice defects.

Here, it should be mentioned that Anderson localization is not the
appearance of a localized eigenmode in a system with one
\emph{isolated} defect, but, as already indicated by the title of the
original work~\cite{Anderson58}, the absence of diffusion in a system
containing \emph{many} defects. The collaborative effect of these
defects leads to a breakdown of the diffusive transport and hence, a
change in the transport characteristic from an algebraic behavior,
typical for a diffusive transport, to an exponential one is a clear
indicator for Anderson localization. Certainly, in the context of
electromagnetic waves also absorption could lead to an exponential
transport behavior and therefore, to make the connection between
Anderson localization and the propagation of electromagnetic waves
through a disordered medium it is important to ensure that the
exponential decay is caused by the disorder and not by absorption
processes.

\begin{figure}[htb]
\epsfig{file=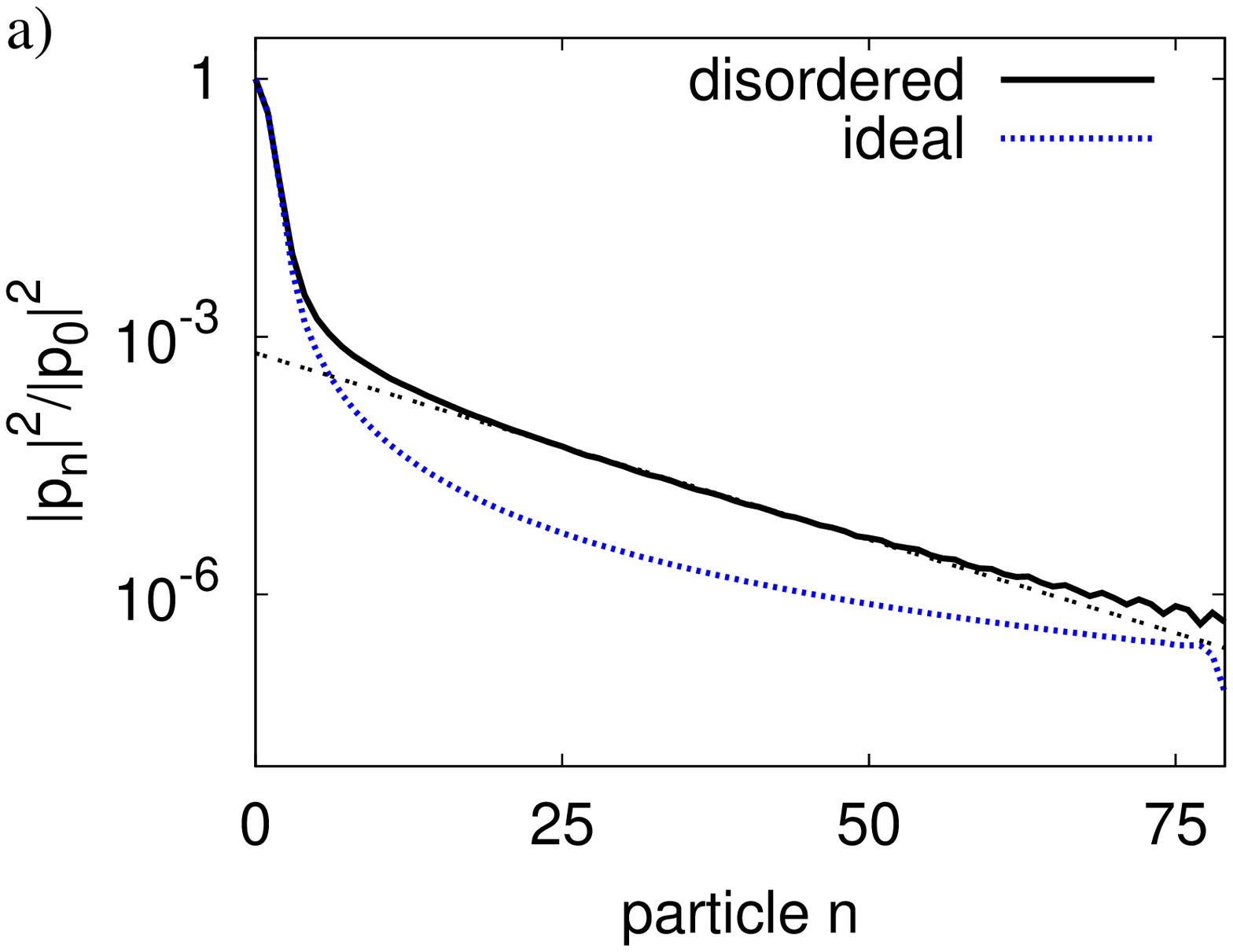, width=0.4\textwidth} \epsfig{file=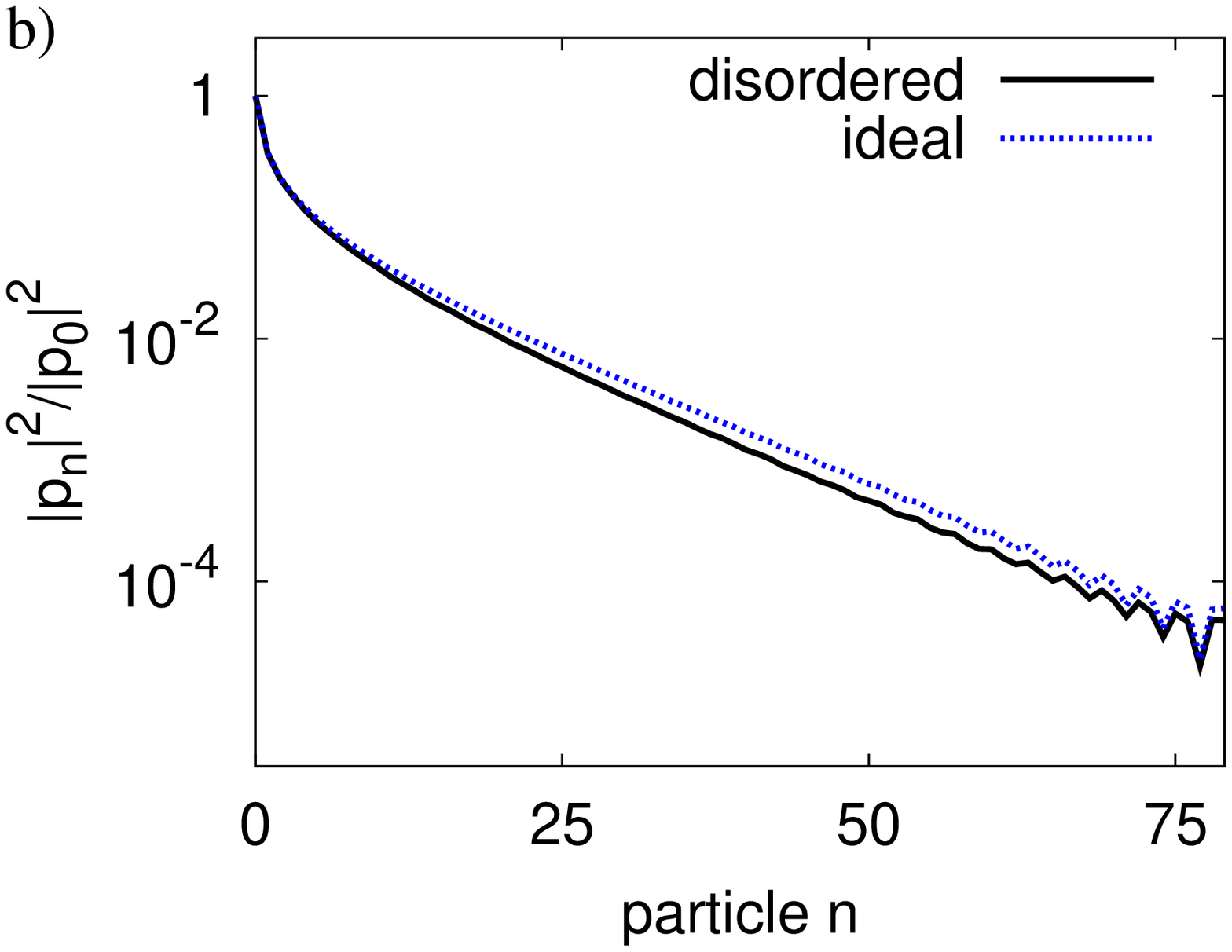,
  width=0.4\textwidth}
\caption{(Color online) Induced dipole moments in chains of 80
  randomly oriented ellipsoids and ideal spheres, respectively, when
  the incident field is present only at the position of the first
  particle, possessing the frequency $\omega=\omega_{spp}$. In a) the
  external field is polarized perpendicular to the chain axis, whereas
  it is polarized parallel to this axis in b). The dashed line in a)
  shows that here the dipole moments decay almost exponentially in the
  transversely excited disordered chain.}
\label{Fig:F_2}
\end{figure}

For studying Anderson localization in chains of nanoparticle, the
transport through 80-element chains of randomly oriented ellipsoids
with normally distributed semi-axes is investigated, assuming a mean
of $a_0$ and a standard deviation of $\sigma a_0$. In
Fig.~\ref{Fig:F_2} the results obtained when only the first particle
is subjected to the incident field is plotted for $\sigma=0.05$, and
compared with the transport through a chain of spheres. Again
the center-to-center distance was $80\,\text{nm}$ and
$a_0=25\,\text{nm}$; here and in the following the results in the
disordered cases are averaged over 1000 realizations.

In the case of transversal excitation shown in Fig.~\ref{Fig:F_2} a)
the disorder obviously causes a substantial \emph{increase} of the
induced dipole moments as compared to the ideal chain; observe the
logarithmic scale. Most significantly, the decay of the moments along
the chain changes from algebraic to nearly exponential when disorder
is introduced, as indicated by the dashed line. This is, as explained
above, a genuine hallmark of disorder-induced localization. In
contrast, for longitudinal excitation the disorder only leads to a
marginal \emph{decrease} of the induced dipole moments.

Crucial for the considerable effect of the disorder in the case of
transversal excitation is that the disorder disturbed the symmetry of
the system. Hence, the effect does not depend on the precise type of
disorder, only the parallelism of the incident field and the induced
moments has to be circumvented. The strong effect induced by this
rotating of the dipoles results from the heavy direction-dependency of
the dipole field (see Eq.~(\ref{Eq:dip})).

\begin{figure}[htb]
\epsfig{file=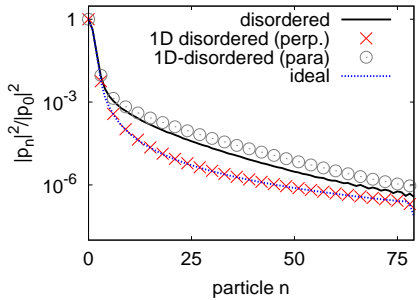, width=0.4\textwidth} 
\caption{(Color online) Induced dipole moments in chains of 80
  spheres. In the ideal case the particles are arranged on a 1D linear
  chain with a center-to-center distance of $80\,\text{nm}$, while the
  disorder is introduced due to normally distributed variations of the
  particles' positions (with a mean of 0 and a standard deviation of
  $5\,\text{nm}$). The positions are varied in all directions
  (\emph{disordered}, see Eq.~(\ref{Eq:dis_pos})), or only in one
  dimension (perpendicular (see Eq.~(\ref{Eq:dis_posx})) or parallel
  (see Eq.~(\ref{Eq:dis_posz})) to the incident field),
  respectively. The incident field is present again only at the
  position of the first particle, possessing the frequency
  $\omega=\omega_{spp}$ and polarized perpendicular to the chain
  axis. For clarity in the 1D cases only the induced moments in every
  third particle are depicted.}
\label{Fig:F_3}
\end{figure}

This is demonstrated in Fig.~\ref{Fig:F_3} where the disorder is not
introduced due to distortions of the elements' shapes, but due to
variations of the positions of the particles. If one assumes that the
chain axis is oriented along the $x$-axis of the coordinate system
used the sites of the ideal chain could be labeled by
\begin{equation}
\vec{r}_n^{ide}=nd\vec{e}_x
\end{equation}
with the center-to-center distance $d$. The disorder is now introduced
due to normally distributed deviations $a_n^{x,y,z}$ from these ideal
positions
\begin{equation}
\label{Eq:dis_pos}
\vec{r}_n^{dis}=nd\vec{e}_x+a_n^x\vec{e}_x+a_n^y\vec{e}_y+a_n^z\vec{e}_z.
\end{equation}
In Fig.~\ref{Fig:F_3} the results for three cases of this type of
disorder are depicted for a transversal excitation. In every case the
mean of $a_n^{x,y,z}$ was zero with a standard deviation of
$5\,\text{nm}$ and a site distance $d$ of $80\,\text{nm}$. This
disorder again leads to an exponential decay as demonstrated by the
curve labeled \emph{disordered} in Fig.~\ref{Fig:F_3}. However, for
this type of disorder it is easily possible to change the disorder
such, that the induced dipole moments and the incident field remain
parallel. Due to the structure of the dipole field (see
Eq.~(\ref{Eq:dip})) a variation of the chain elements' position
perpendicular to the incident field does not affect the orientation of
the induced dipoles. To clarify this, assume a polarization of the
incident field in $z$-direction with the chain-axis again in the
$x$-direction. Since, for a perfect sphere, the induced dipole moment
and the field at the position of the sphere are parallel, all induced
moments in the chain and the incident field are parallel if the fields
at all chain sites are parallel to the external field. If now the
positions of the particles are varied only perpendicular to the
polarization of the external field, for example only in $x$-direction,
\begin{equation} 
\label{Eq:dis_posx}
\vec{r}_n^{dis}=nd\vec{e}_x+a_n^x\vec{e}_x,
\end{equation}
for all $n$, the dot product of $\vec{r}_n^{dis}$ with the
polarization of the incident field ($\propto \vec{e}_z$) vanishes, and
the component of the field along $\vec{r}$ in Eq.~(\ref{Eq:dip}) is
zero. Therefore, the fields produced by the dipole moment in
$z$-direction in the first particle, caused by the external field, are
also polarized in $z$-direction at all positions $\vec{r}_n^{dis}$
(see Eq.~(\ref{Eq:dip})).  Since, similar arguments hold for the fields
produced by the dipole moments of all other chain members, all induced
dipole moment are parallel to the incident field. Thus, a disorder
described by Eq.~(\ref{Eq:dis_posx}) changes the induced moments
only marginally, as demonstrated by the red crosses in
Fig.~\ref{Fig:F_3}. The same qualitative result is obtained if instead
of non-zero $a_n^x$, non-zero $a_n^y$ or even non-zero $a_n^x$ and
non-zero $a_n^y$ are used. But, if the positions are varied in
$z$-direction
\begin{equation} 
\label{Eq:dis_posz}
\vec{r}_n^{dis}=nd\vec{e}_x+a_n^z\vec{e}_z,
\end{equation}
the dot products of $\vec{r}_n^{dis}$ with the polarization of the
incident field vanish no longer and hence, in this case the induced
dipole moments and the external field are in general not parallel. This results
in a qualitative change of the transport behavior as exemplified  by
the gray circles in Fig.~\ref{Fig:F_3}.

\begin{figure}[htb]
\epsfig{file=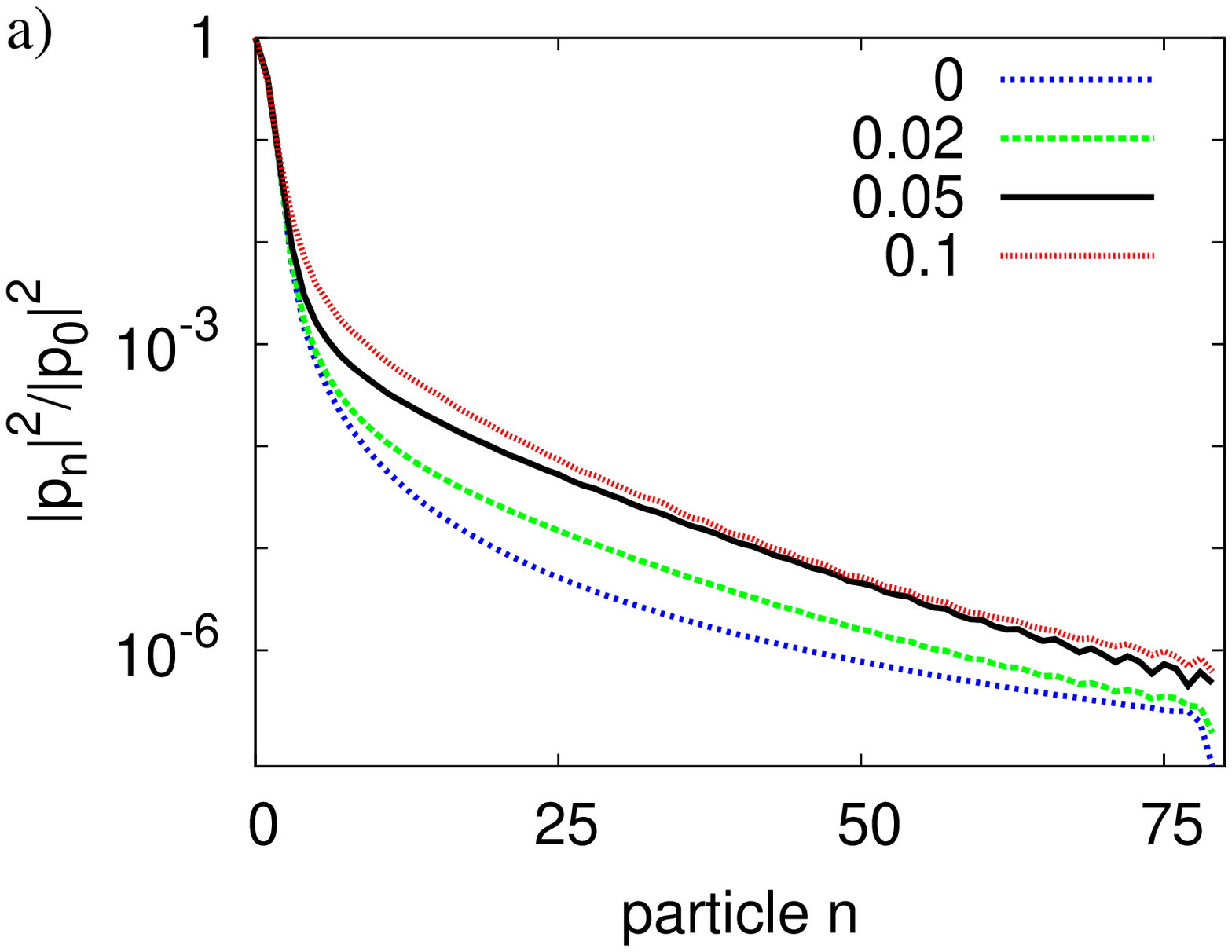, width=0.4\textwidth}
\epsfig{file=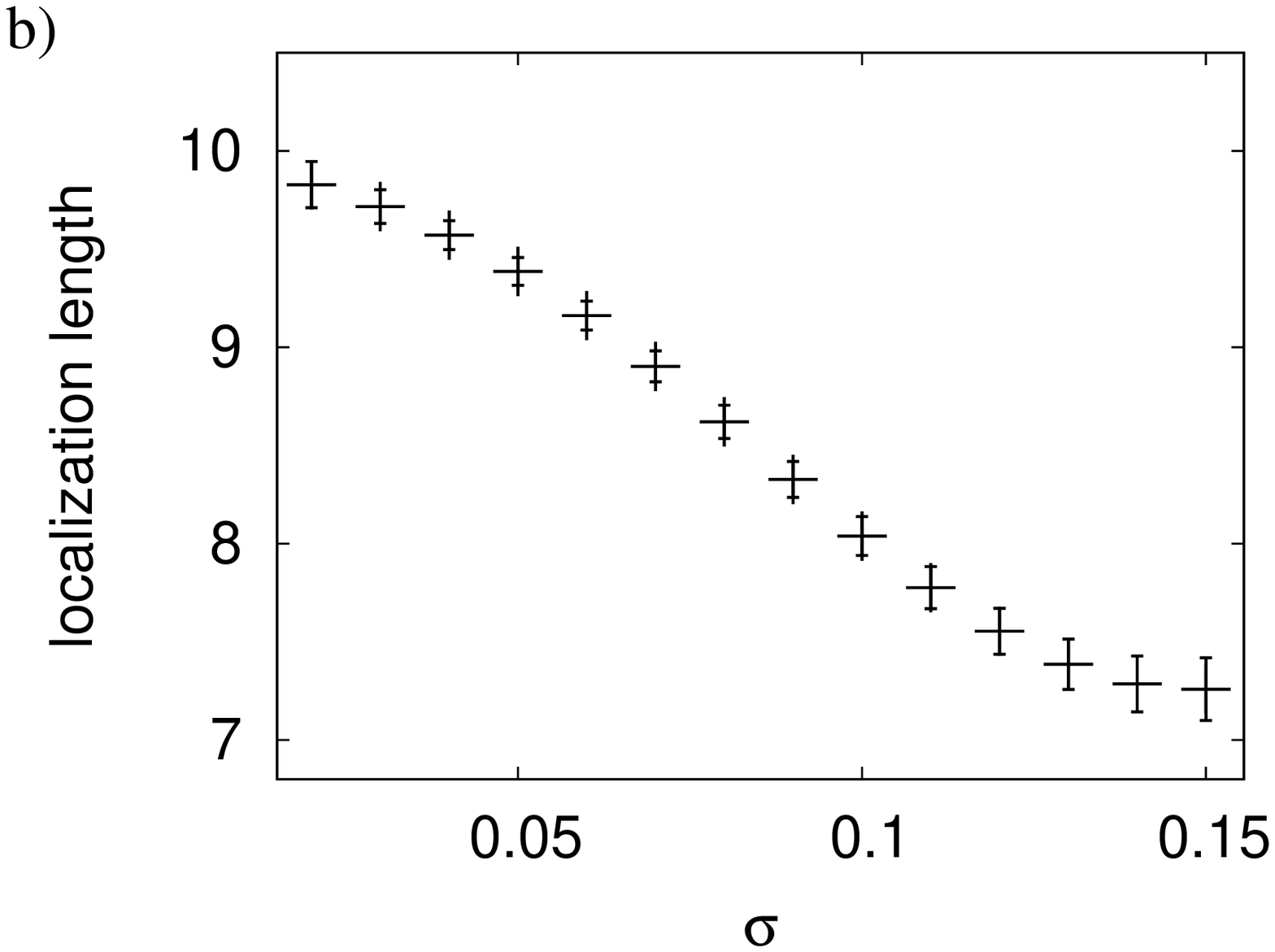, width=0.4\textwidth}
\caption{(Color online) Dependence of the transport characteristics on
  the strength $\sigma$ of the distortion of the spheres, for
  transversal excitation. Panel a) shows the induced dipole moments
  for different values of $\sigma$, while panel b) depicts the
  dependence of the localization length $b$ (in numbers of particles,
  see Eq.~(\ref{Eq:p})) on $\sigma$, with numerical errors as
  indicated by the symbols. In all cases the frequency of excitation
  was $\omega=\omega_{spp}$.}
\label{Fig:F_4}
\end{figure}

Naturally, the transport characteristics of the disordered chain
depend on the strength of the particular kind of disorder, in addition
to the center-to-center distances and to the optical properties of the
nanoparticles. For studying the dependence of the distortion strength
again chains of randomly oriented ellipsoids with normally distributed
semi-axes with a mean of $a_0$ and a standard deviation of $\sigma a_0$ are
used. In Fig.~\ref{Fig:F_4} the transport is depicted for different
strengths $\sigma a_0$, for the interesting case of transversal
excitation. Here, the cross-over from an algebraic to the exponential
decay is visible. In order to extract the dependence of the
localization length $b$ on the distortion strength $\sigma$, the
dipole moments in the middle part of the chain have been fitted to the
expression
\begin{equation}
\label{Eq:p}
|p_n|^2=a\exp(-n/b).
\end{equation}
The results are shown in Fig.~\ref{Fig:F_4}b). As expected, an
increase of the disorder strength causes a decrease of the
localization length, and hence a ``stronger'' localization. Since
Anderson localization is caused by multiple scattered waves, the
variation of the individual chain members should not be too
large. After increasing the value of $\sigma$ to 0.2, for instance,
one does no longer find an exponential decay of the induced dipole
moments along the chain. For the randomly oriented ellipsoids this is
due to the fact that the chain is excited with the resonance frequency
of an undistorted chain member and that the distortion of the
particles also affects the resonance frequencies; a too large
distortion decreases significantly the scattering cross-section of the
particle at the frequency used and thus decreases the effect of
multiple scattered waves.

\begin{figure}[htb]
\epsfig{file=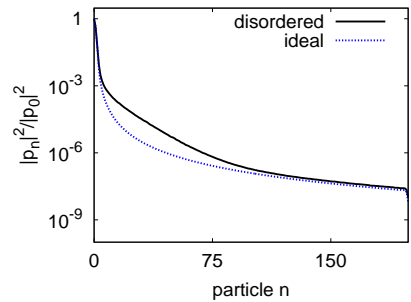, width=0.4\textwidth}
\caption{(Color online) As Fig.~\ref{Fig:F_2}a), but for a chain of
  200 particles. Note that the induced dipole moments in the
  disordered chain do not drop below those of the ideal one.}
\label{Fig:F_5}
\end{figure}

Perfect exponential localization would imply that for longer chains
the induced dipole moments can drop substantially below those in
ideal chains. Somewhat surprisingly, this does not seem to occur in
the system studied here. Instead, in longer chains there appears to be
a second cross-over after which the response of the disordered chain
again approaches that of the ideal one. This is exemplified in
Fig.~\ref{Fig:F_5} for a chain of 200 particles. Thus, there are at
least two characteristic length scales: That which marks the onset of
the fully developed exponential decay, and that which marks the return
to the behavior of the ideal chain.

\section{Conclusions}
\label{Sec:Conclusions}
In conclusion, chains of randomly or even deliberately distorted
nanoparticles constitute novel experimentally accessible and
theoretically tractable systems for studying localization effects of
electromagnetic waves at optical frequencies. Utilizing these
localization effects in combination with specific types of the
engineered disorder offers the prospect of obtaining particularly high
field enhancement for application purposes.

\begin{acknowledgments}
The author thanks D. Grieser, M. Holthaus, and H. Uecker for useful
discussions. 
\end{acknowledgments}


\begin{thebibliography}{99}
\bibitem{footnote2} For determining the eigenmodes, the set
  of equations~(\ref{eq:polar}) can be recast into matrix form
  $M(\omega)\vec{p}=\vec{E_{ext}}$ with $\vec{E_{ext}}$ denoting the
  external electric fields present at the positions of the
  nanoparticles. The frequencies of the eigenmodes are given by
  the complex roots of $\det(M)=0$, and the corresponding eigenmodes by
  the nontrivial solutions of $M\vec{p}=0$~\cite{Weber04}.
\end{thebibliography}
\end{document}